\documentclass[prb,twocolumn,superscriptaddress]{revtex4-1}

\usepackage{amsmath}
\usepackage{amssymb}
\usepackage{xspace}
\usepackage{graphicx}
\usepackage{grffile}
\usepackage{nicefrac}
\usepackage{color}

\graphicspath{{figs/}}

\begin{document}

\title{Electron correlation and magnetism at the LaAlO$_3$/SrTiO$_3$ interface:\\
A DFT+DMFT investigation}

\author{Frank Lechermann}
\affiliation{I. Institut f{\"u}r Theoretische Physik, Universit{\"a}t Hamburg, 
D-20355 Hamburg, Germany}
\author{Lewin Boehnke}
\affiliation{I. Institut f{\"u}r Theoretische Physik,
Universit{\"a}t Hamburg, D-20355 Hamburg, Germany}
\author{Daniel Grieger}
\affiliation{International School for Advanced Studies (SISSA),
Via Bonomea 265, I-34136 Trieste, Italy}
\author{Christoph Piefke}
\affiliation{I. Institut f{\"u}r Theoretische Physik,
Universit{\"a}t Hamburg, D-20355 Hamburg, Germany}

\begin{abstract}
We shed light on the interplay between structure and many-body effects
relevant for itinerant ferromagnetism in LaAlO$_3$/SrTiO$_3$ heterostructures.
The realistic correlated electronic structure is studied by means of the (spin-polarized) 
charge self-consistent combination of density functional theory (DFT) with 
dynamical mean-field theory (DMFT) beyond the realm of static correlation 
effects. Though many-body behavior is also active in the defect-free interface, 
a ferromagnetic instability occurs only with oxygen vacancies. A minimal Ti 
two-orbital $e_g$-$t_{2g}$ description for the correlated subspace is derived. 
Magnetic order affected by quantum fluctuations builds up from effective double 
exchange between modified nearly-localized $e_g$ and mobile $xy$ electrons.
\end{abstract}


\maketitle

\section{Introduction}
The interface physics emerging from the combination of LaAlO$_3$ (LAO) and 
SrTiO$_3$ (STO) bulk band insulators highlights the prominent
research on oxide heterostructures (see e.g. Refs.~\onlinecite{zub11,hwa12} for a review). 
Crystal growth along (001) results in a metallic two-dimensional electron system 
(2DES) for an AlO$_2$-LaO-TiO$_2$ (n-type) boundary region.~\cite{oht04,thi06}.
Since that stacking formally leads to a diverging electrostatic 
potential with increasing thickness, electronic reconstruction to avoid the
polar catastrophe~\cite{nak06} is a possible explanation for the 2DES. 
Intrinsic doping via oxygen vacancies~\cite{kal07,sie07} may also play a role,
as demonstrated in view of itinerancy at vacuum-cleaved STO surfaces.~\cite{san11}
Surprisingly, ferromagnetic (FM) 
and superconducting order~\cite{rey07,bri07,li11,ari11} may be stabilized in 
LAO/STO interfaces. Ferromagnetism is either bound to
stoichiometry~\cite{lee13} or associated with defects,~\cite{sal13} but
in any case, electron correlations are assumed to be important. Coexistence of 
itinerant and localized electrons is suggested from scanning-tunneling 
spectroscopy,~\cite{bre10,ris12} anisotropic magnetoresistance, anomalous 
Hall-effect measurements,~\cite{jos13} resonant soft-x-ray scattering~\cite{par13} 
and photoemission.~\cite{san11,mee11,ber13}

Numerous theoretical works address the LAO/STO interface 2DES, ranging from 
model-Hamiltonian studies~\cite{che13_2,kha13,ban13,zho13,pav13,lin13,ruh13} 
to density functional 
theory (DFT) (+Hubbard $U$) investigations.~\cite{pen06,rey07,pav12,kha13,zho13}
Though agreement exists about focusing on the Ti$(3d)$ shell, differences  
concerning crucial subshell states and the relevance of crystal defects persist. 
Because of low Ti filling at stoichiometry, there sole {\sl local} Coulomb 
correlations are not expected sufficient for phenomenology-relevant 
many-body physics.~\cite{ban13,che13_2} Inspired by DFT+U calculations
with interface oxygen vacancies,~\cite{pav12} Pavlenko {\sl et al.} used  
a basic $e_g$-$t_{2g}$ Hubbard model~\cite{pav13} to consider the itinerant/localized 
signature in Hartree-Fock approximation. Other many-body modellings focus on 
defect-driven electronic states within an impurity/Kondo scope.~\cite{lin13,ruh13}
Yet a first-principles many-body revelation connecting (defect) structure,
itinerant orbital-spin state and ferromagnetism is still lacking.

We here remove this deficiency and reveal the subtle interplay between defect
occurrence and many-body effects in supporting metallic ferromagnetism at 
realistic LAO/STO interfaces, based on a charge self-consistent DFT + dynamical 
mean-field theory (DMFT) study. While local self-energies do not promote FM order in
the defect-free case, with oxygen vacancies effective double exchange governed by the 
Hund exchange $J_{\rm H}$ near quarter filling drives an intricate FM phase subject to quantum 
fluctuations. The connected key correlated subspace is readily derived as being composed of
vacancy-induced almost-localized tailored $e_g$ and quasi-itinerant $t_{2g}(xy)$ states.

\section{Crystal Structure and Theoretical Framework}
\begin{figure}[t]
(a)\includegraphics*[height=6.5cm]{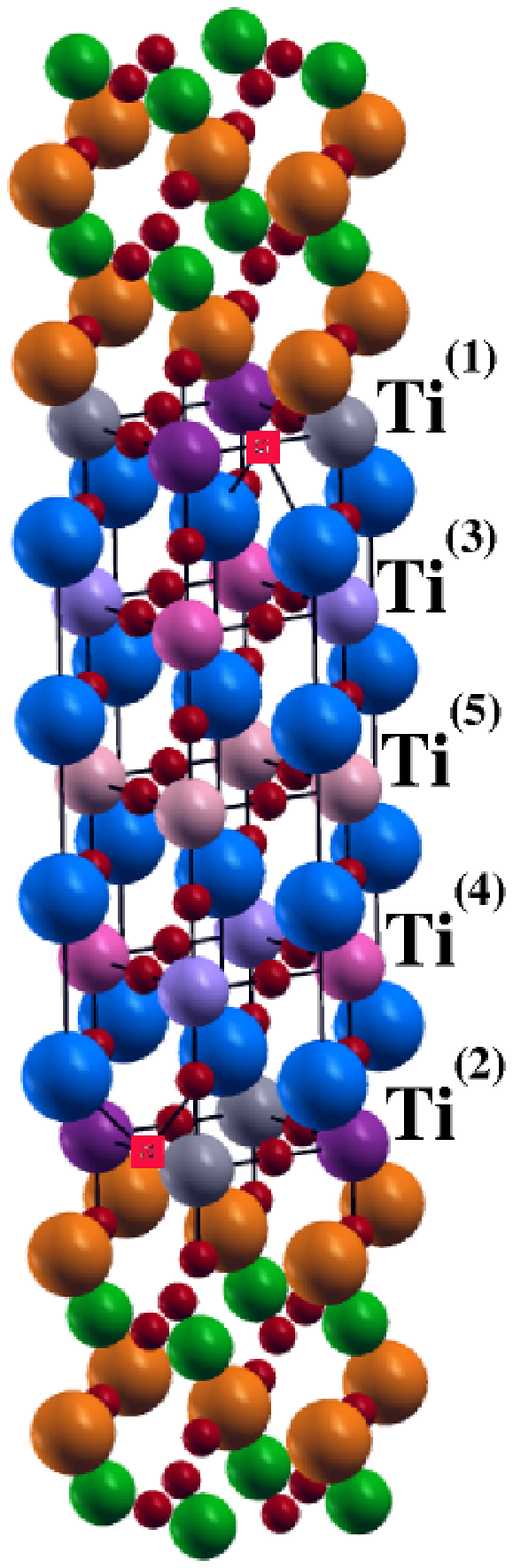}
(b)\includegraphics*[width=3.25cm]{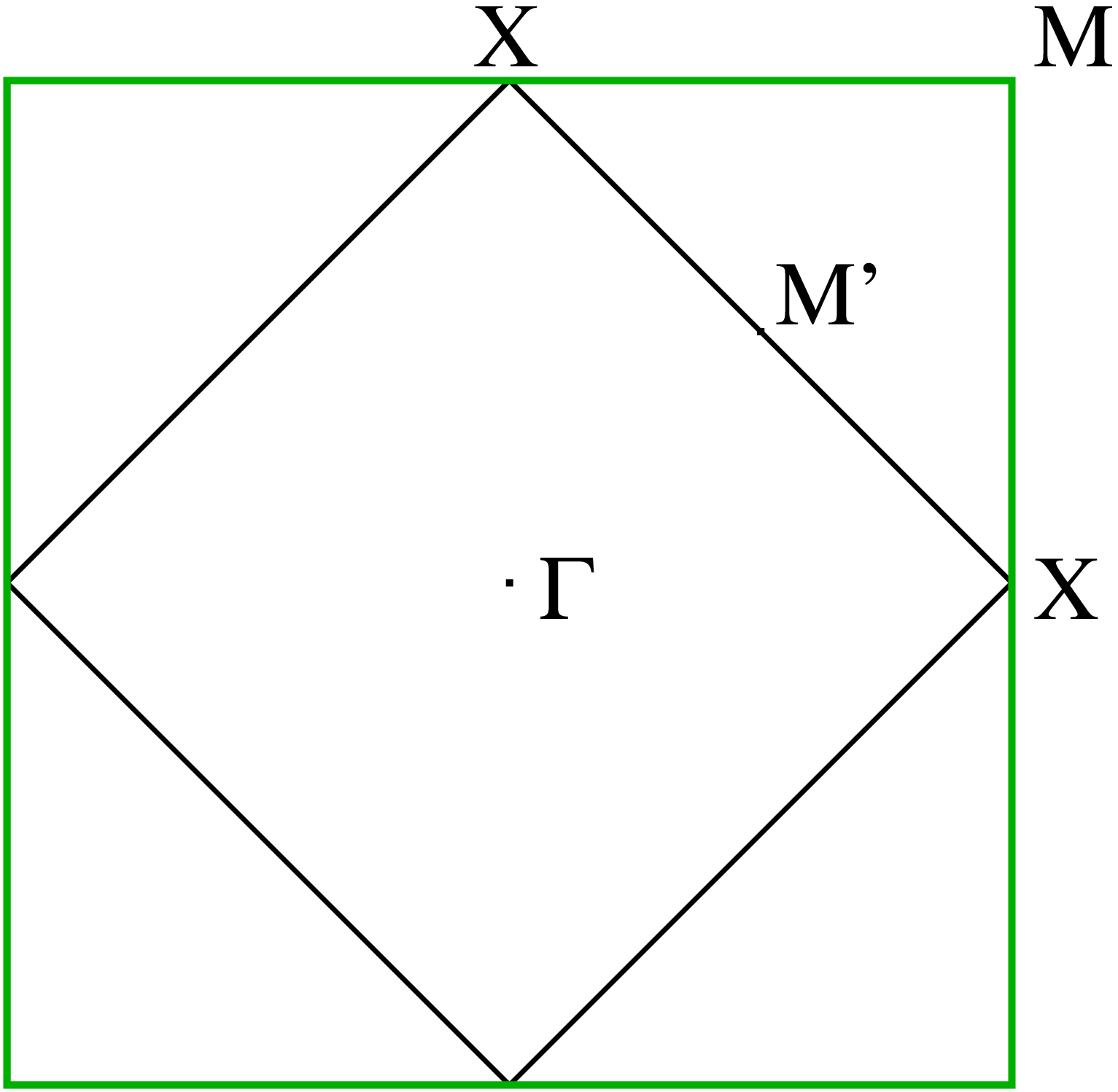}
\caption{(color online) (a) Ideal LAO/STO structure:
La (large orange/lightgrey), Sr (large blue/grey), Al (small green/grey), 
O (small red/dark) ions and symmetry-inequivalent Ti ions. 
Small squares mark O vacancies. (b) $(k_x,k_y)$ Brillouin zone for the here 
used $\sqrt{2}$$\times$$\sqrt{2}$ structuring with two inplane Ti atoms (black). 
The basic 1$\times$1 BZ is given in green (grey). The high-symmetry points
read $\Gamma$=(0,0,0), $M'$=$(\frac{\pi}{4a},\frac{\pi}{4a},0)$, 
$M$=$(\frac{\pi}{2a},\frac{\pi}{2a},0)$, $X$=$(0,\frac{\pi}{2a},0)$.\label{fig1:struclda}}
\end{figure}

A superlattice with four LaO(SrO) layers, each having a $\sqrt{2}$$\times$$\sqrt{2}$
inplane unit cell, models the n-type interface (see Fig.~\ref{fig1:struclda}a). 
This is just above the minimal LAO thickness-limit for the onset of
FM order.~\cite{kal12} The supercell is based on 80 atomic sites. We utilize the STO 
lattice constant $a$=3.905\AA$\,$ in the planes perpendicular to the $c$ 
axis, while the $c/a$ ratio is optimized to $c/a$=0.986 for the whole cell from total-energy 
minimization. In general the local $c/a$ is somewhat larger unity in the STO part, whereas its 
below unity in the LAO part.

The impact of defects is examined from two interface limits. A defect-free (DF) case and 
an oxygen-vacancy-hosting (VH) one with high defect concentration. One O vacancy per 
interface is placed in the boundary TiO$_2$ layer.~\cite{sie07,she12,liu13} 
Asymmetric vacancy positions are chosen to reduce defect coherency. The 25$\%$ vacancies per 
interface TiO$_2$ exceed usual experimental magnitudes, but the modelling is geared to grasp 
the key doping effect. Three inequivalent TiO$_2$ layers are identified. 
One may inplane symmetrize each inplane Ti pair, but Ti$^{(1)}$, Ti$^{(2)}$ as well as 
Ti$^{(3)}$, Ti$^{(4)}$ are treated inequivalent to allow for intra-layer ordering tendencies.

Local structural relaxation allowed for the defect-free  and the vacancy-hosting 
structure is performed by minimizing the atomic forces within the 
local density approximation (LDA). A mixed-basis pseudopotential scheme is utilized
for the LDA calculations. Using localized functions for Ti$(3d)$ and O$(2s2p)$ in 
the Kohn-Sham basis renders it possible to reduce the plane-wave cutoff energy to 
moderate 13 Ryd for the large cell. Up to an 9$\times$9$\times$3 tailoring is 
used for the $k$-point mesh in reciprocal space. The two-dimensional Brillouin zone (BZ) of 
our enlarged inplane unit cell is shown in Fig.~\ref{fig1:struclda}b. Note
that the $M'$ point discussed here is located half the distance to $\Gamma$ compared to the 
original $M$ of the basic square BZ for the single-atom square/cubic structure.

Charge self-consistent DFT+DMFT~\cite{sav01,pou07,gri12} is build up on the derived orbital 
projection from 30 Kohn-Sham bands following the Ti$(3d)$-like manifold's lower band 
edge.~\cite{ama08} At each Ti site a rotational-invariant Slater-Kanamori parametrization
of the multi-orbital Hubbard Hamiltonian, i.e. Hubbard $U$ and Hund's exchange $J_H$, is 
applied. The local interacting problem on each individual Ti site is 
described by the many-body Hamiltonian
\begin{eqnarray}
{\cal H}&=& 
U\sum_{m} n_{m\uparrow}n_{m\downarrow}+\frac{1}{2}\sum \limits _{m\ne m',\sigma}
\Big\{U' \, n_{m \sigma} n_{m' \bar \sigma}+\nonumber\\
&&+\,U'' \,n_{m \sigma}n_{m' \sigma}+
\,J_H\left(c^\dagger_{m \sigma} c^\dagger_{m' \bar\sigma} 
c^{\hfill}_{m \bar \sigma} c^{\hfill}_{m' \sigma}+\right.\\   
&&\left.+\,c^\dagger_{m \sigma} c^\dagger_{m \bar \sigma}  
 c^{\hfill}_{m' \bar \sigma} 
c^{\hfill}_{m' \sigma}\right)\Big\}\;,\nonumber      
\label{eq:hubbardham}          
\end{eqnarray} 
with orbitals $m$,$m'$=1,2,(3), spin projection $\sigma$=$\uparrow,\downarrow$ and
$n$=$c^\dagger c$. Inter-orbital density-density terms scale  
with $U'$=$U$$-$$2J_H$ and $U''$=$U$$-$$3J_H$. Computing $U$, $J_H$ from first 
principles is tough for the complicated crystal structures. Therefore we choose suitable
values to access key LAO/STO correlation physics, whereby the Coulomb 
interactions are approximated site independent. The values $U$$\sim$5 eV, $J_H$$\sim$0.7 eV
are a proper choice~\cite{miz95} for three-orbital models of {\sl bulk} titanates. 
Because of the weakly correlated LaAlO$_3$ part and the use of orbital-reduced schemes, 
we lower to moderate $U$=3.5(2.5) eV and $J_H$=0.5 eV for the three(two)-orbital 
case~\cite{bre10,pav13} (see Sec.~\ref{corrspace}). Five inequivalent impurity 
problems are solved at each step of full DFT+DMFT~\cite{gri12} in the supercell, utilizing 
the hybridization-expansion version of continuous-time quantum Monte 
Carlo.~\cite{rub05,wer06,triqs_code,boe11} Computations are performed at temperature 
$T$=145.1K ($\beta$$\equiv$$1/T$=80eV$^{-1}$), if not otherwise stated.
A double-counting correction of the fully-localized form~\cite{sol94} is applied. The 
maximum-entropy method is used for the analytical continuation from Matsubara space to
obtain the spectral data.

In the spin-polarized ferromagnetic case, the Kohn-Sham part is handeled within the 
local spin density approximation (LSDA) and the complete {\sl spin-resolved} charge density 
is converged. Only for a rough estimation of the Curie temperature $T_{\rm c}$ for the 
VH structural case the {\sl spin-averaged} charge self-consistency cycle with 
however of course {\sl spin-resolved} DMFT self-energies is utilized.

\section{LDA band structure and density of states}
\begin{figure}[t]
\hspace*{-0.4cm}
\parbox[c]{1.85cm}{\includegraphics*[width=2.1cm]{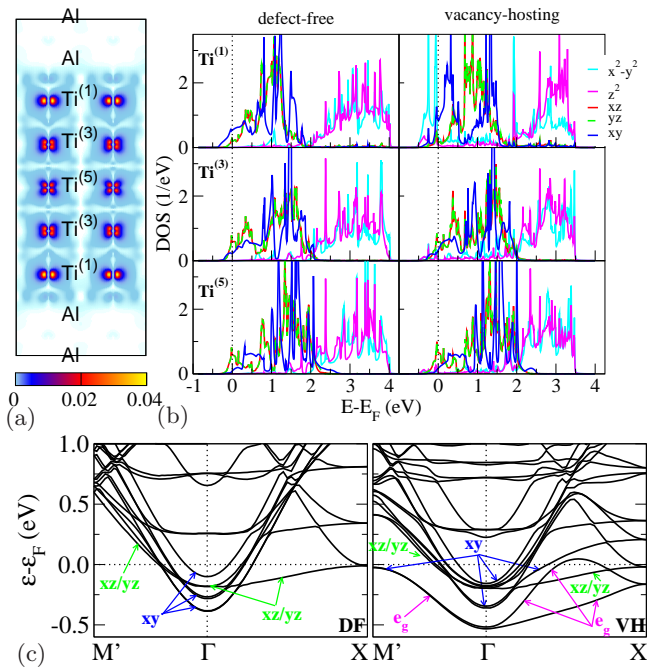}\\[-0.15cm]
\raggedright(a)}
\hspace*{-0.05cm}
\parbox[c]{6.4cm}{(b)\hspace*{-0.45cm}\includegraphics*[width=6.5cm]{DOS-tmo.eps}}\\
(c)\hspace*{-0.4cm}\includegraphics*[width=8.4cm]{BANDS-tmo.eps}
\caption{(color online) LDA data for LAO/STO. (a) Valence charge density of the occupied
low-energy manifold in the DF structure. (b) Orbital-resolved DOS of projected Ti$(3d)$. 
Ti$^{(2)}$(Ti$^{(4)}$) data are identical to Ti$^{(1)}$(Ti$^{(3)}$) data (see text).
(c) Band structures along high-symmetry lines.\label{fig2:dosbands}}
\end{figure}

Metallicity is obtained for both structure types from DFT in LDA 
(see Fig.~\ref{fig2:dosbands}). At stoichiometry, 
two electrons occupy the dominant Ti$(3d)$ low-energy manifold, matching the number
for polar-catastrophe avoidance, which predicts Ti$^{3.5+}$O$_2^{4-}$ at the 
interface.~\cite{nak06} Figure~\ref{fig2:dosbands}a shows that the two electrons which 
form the 2DES are confined to the STO part, with dominant localization near Ti$^{(12)}$,
 i.e. directly at the boundary towards the LAO part.

Six electrons settle in the Ti$(3d)$-like manifold with vacancies. 
There the occupied bandwidth amounts to $\sim$0.5 eV, with two additionally 
filled bands along $\Gamma$$-$$M'$ in the Brillouin zone. An orbital-character analysis 
renders these two bands $e_g$-like, yet with sizeable $xy$ weight close to $M'$. For the 
$t_{2g}$ part the $xz/yz$-like bands have in general larger LDA effective mass 
than $xy$-like. The new bands are indeed associated with Ti$(e_g)$ states 
liberated from O$(2p)$, as also evident from the local Ti($3d$)-resolved density of 
states (DOS) (cf. Fig.~\ref{fig2:dosbands}c). In real space, the LDA layer
bonding charge density (Fig.~\ref{fig3:charge}a,b) exhibits directly $e_g$-like 
filling of Ti$^{(1)}$, Ti$^{(2)}$ (=Ti$^{(12)}$) in addition to $xy$. The former 
share is absent in the far-from-interface Ti$^{(5)}$O$_2$ layer. 

Spin-polarized LDA does not support ferromagnetism with a sizable supercell 
net moment ($m_{\rm tot}$$\le$$0.01\mu_{\rm B}$) for either structure type. Thats 
corroborated by the failing Stoner criterion $IN(\varepsilon_{\rm F})$$>$1 in the DFT 
context~\cite{gun76} which should signal an itinerant FM state for Stoner parameter 
$I$ and DOS $N(\varepsilon_{\rm F})$ per Ti spin at the Fermi level. 
The value of $I$ for elemental Ti is about 0.6 eV~\cite{bak93} and a computation 
directly for LAO/STO by Janicka {\sl et al.} yields 0.725 eV.~\cite{jan08} Our obtained 
$N(\varepsilon_{\rm F})$=0.63 (0.77) eV$^{-1}$ per spin for the DF (VH) interface 
are thus not sufficient to allow for conventional band magnetism.

\section{Correlated Subspace\label{corrspace}}
Correlations are important and we here treat those beyond existing static 
DFT+U studies.~\cite{pen06,pav12} First we elaborate on the minimal correlated 
subspace~\cite{ani05,ama08,hau10} for the LAO/STO 2DES. An $e_g$-$t_{2g}$ modelling is
advocated from the LDA findings. However a complete correlated $3d$-shell 
five-orbital description of Ti$^{(1-5)}$ is very challenging. Instead we devise a 
simpler approach guided by low-energy states. 

\begin{figure}[t]
\parbox[c]{3.75cm}{
\hspace*{-0.2cm}(a)\includegraphics*[height=3.2cm]{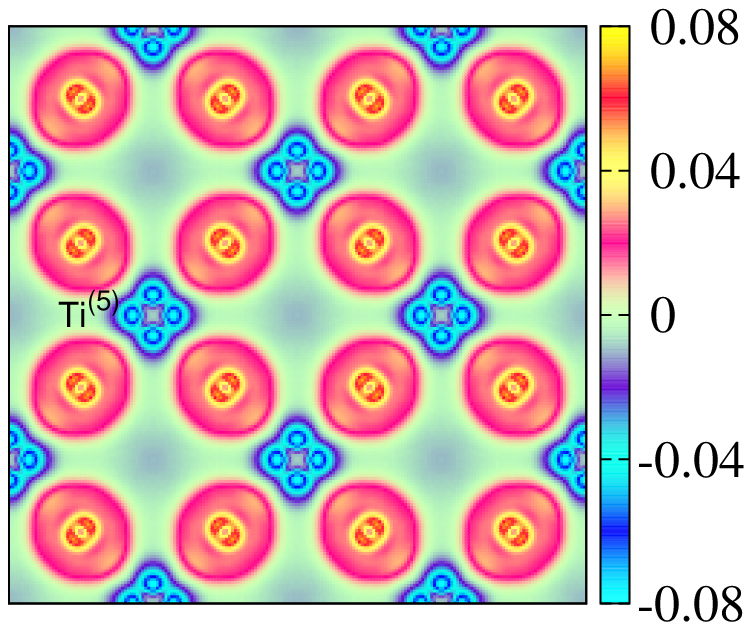}\\
\hspace*{-0.2cm}(b)\includegraphics*[height=3.2cm]{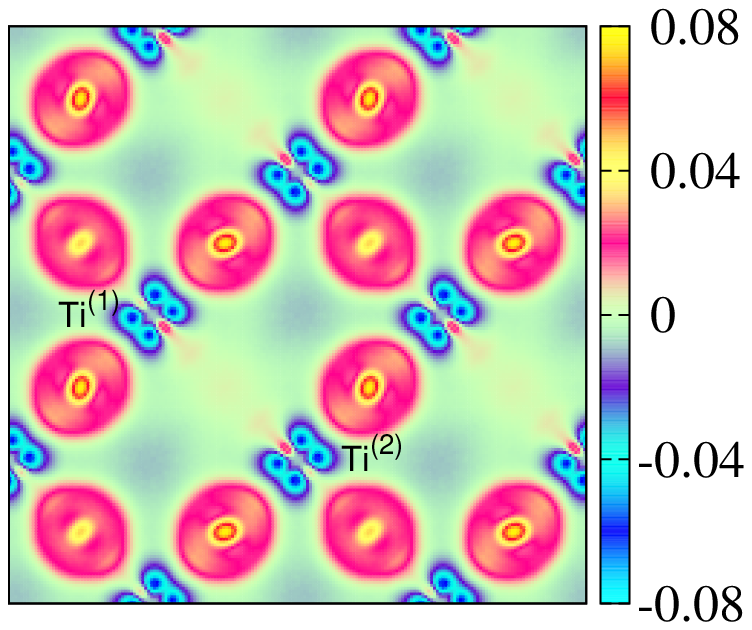}}\hspace*{0.4cm}
\parbox[c]{4cm}{\includegraphics*[height=6.2cm]{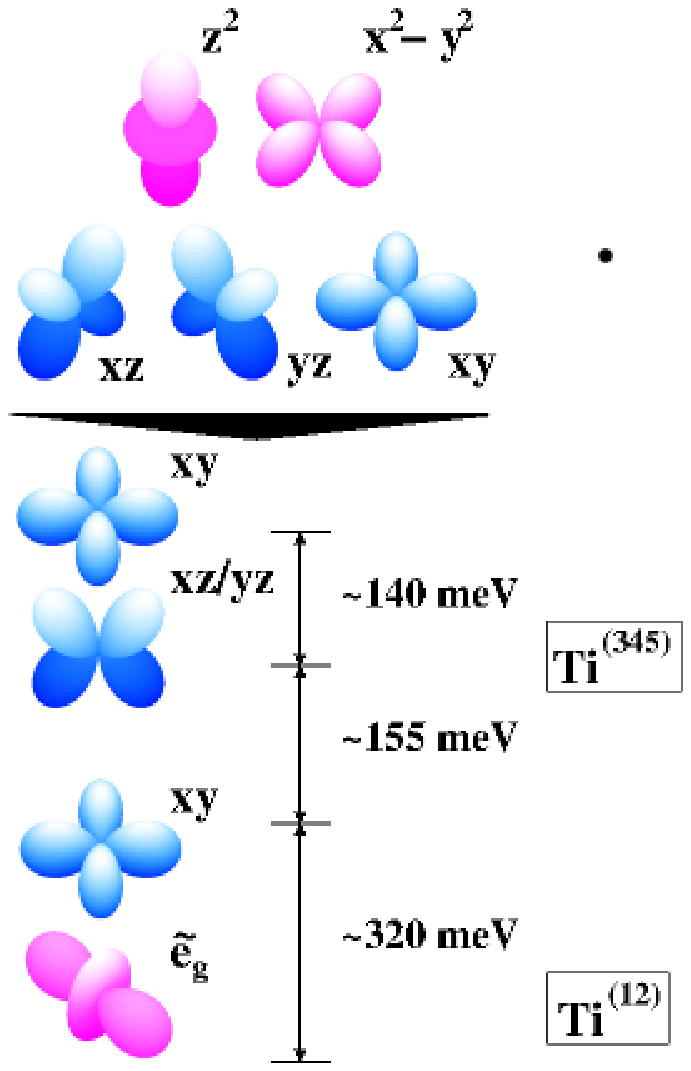}\\
\raggedright{(c)}}
\caption{(color online) (a,b) LDA bonding charge density
$\rho_{\rm b}^{\rm(LDA)}$=$\rho_{\rm tot}^{\rm(LDA)}$$-$$\rho_{\rm atomic}^{\rm(LDA)}$ 
for VH 4$\times$2 LAO/STO. (a) Ti$^{(5)}$O$_2$ layer and (b) interface 
Ti$^{(12)}$O$_2$. (c) Tailored Ti orbital basis derived from original $3d$ states 
along with average LDA level splittings. The $\tilde{e_g}$ orbital reads
$|\tilde{e_g}\rangle$$\,\sim\,$$0.55|z^2\rangle\pm 0.84|x^2$$-$$y^2\rangle$.
\label{fig3:charge}}
\end{figure}

For the DF interface an $t_{2g}$-based 
three-orbital correlated subspace for all Ti ions is adequate 
(see Fig.~\ref{fig2:dosbands}b). The VH structure with its sharing $e_g$-$t_{2g}$ 
spectral structure close to $\varepsilon_{\rm F}$ asks for more. Yet we only treat 
$e_g$ character directly at the interface, i.e. only for Ti$^{(12)}$. Albeit there is 
some $e_g$ leaking into the Ti$^{(34)}$O$_2$ layer, an {\sl unoccupied} low-energy 
$e_g$ part is missing. Small weight without fluctuations to free states may be neglected. 
It follows ($e_g$,$t_{2g}$) for Ti$^{(12)}$ and $t_{2g}$ for Ti$^{(34)}$,  
Ti$^{(5)}$ (=Ti$^{(345)}$). Abandoning five-orbital schemes, a three-orbital construct 
seems proximate, since only $xy$ is $t_{2g}$-sizeable on Ti$^{(12)}$. But a 
three-orbital $z^2$,$x^2$-$y^2$,$xy$ projection of the low-energy LDA bands yields 
strong hybridization between the two $e_g$ orbitals. This was already suggested from the 
directed $\rho_{\rm b}(e_g)$ weight in Fig.~\ref{fig3:charge}b. Diagonalizing the 
local-projected  ($z^2$,$x^2$-$y^2$) Hamilton matrix reduces the $e_g$ problem to a single
relevant $\tilde{e_g}$ orbital on each Ti$^{(12)}$ ion (see Fig.~\ref{fig3:charge}c).
We retrieve a two-orbital ($\tilde{e_g}$,$xy$) correlated subspace in the 
interface TiO$_2$ layer, similar to what was heuristically used in Ref.~\onlinecite{pav13}. 
A respective two-orbital subspace is also sufficient on the remaining Ti ions due to the alike 
behavior of $xz$ and $yz$. Thus a symmetrized $xz/yz$ orbital projection together with 
$xy$ is constructed for Ti$^{(345)}$. Note that in principle a
combined two-orbital Ti$^{(12)}$ and three-orbital Ti$^{(345)}$ treatment (with then 
site-dependent Hubbard $U$) for the VH case would also be possible within 
our DFT+DMFT coding. However since we do not expect significant $xz$ vs. $yz$ polarization 
we choose the symmetrized $xz/yz$ description on Ti$^{(345)}$. 

\section{Many-Body effects}
\begin{figure}[t]
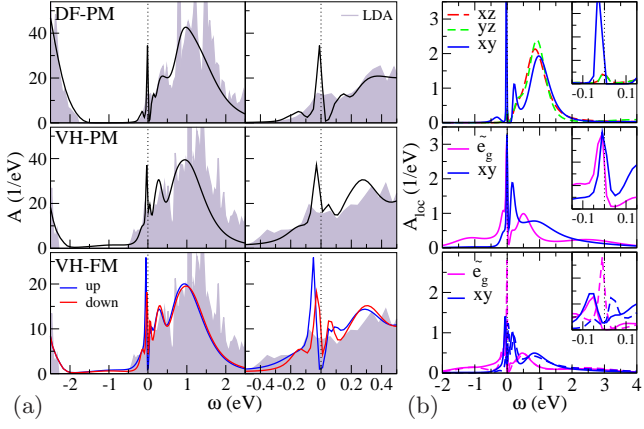

\begin{center}
(a)\hspace*{-0.5cm}\includegraphics*[height=5.5cm]{totspec.eps}
(b)\hspace*{-0.5cm}\includegraphics*[height=5.5cm]{locspec.eps}
\end{center}
\caption{(color online) $k$-integrated DFT+DMFT spectral data. 
(a) Total spectrum including 30 states above O$(2p)$-Ti($3d)$ gap
(right part: blow up around $\varepsilon_{\rm F}$).
(b) Ti$^{(1)}$ impurity defect-free (top) and vacancy-hosting 
PM (middle), FM (bottom, spin-down: dashed lines) data.
\label{fig4:spectrum}}
\end{figure}

Though weakly filled (see Tab.~\ref{tab:fillings}), the stoichiometric DF interface is  
susceptible to Coulomb correlations within DFT+DMFT. Figures~\ref{fig4:spectrum}a,b show 
a sharp quasiparticle (QP) peak and small-scale spectral-weight transfer  
dominantly for $xy$ which is singled out by orbital polarization. By comparison, 
the VH paramagnetic (PM) energy spectrum displays still richer correlation signatures.
Apart from substantial low-energy renormalization Fig.~\ref{fig4:spectrum}a reveals 
a shallow lower Hubbard band. This incoherent excitation is dominantly from the 
$e_g$-like state, as verified by the Ti$^{(1)}$ impurity spectral function 
(cf.~Fig.~\ref{fig4:spectrum}b). The $\tilde{e_g}$ peak position at $\sim$$-1.2$ eV is
in excellent agreement with photoemission data.~\cite{mee11,ber13} A QP peak in favor of $xy$ 
but additional $\tilde{e_g}$ weight completes the itinerancy/localization dichotomy 
originating within the Ti$^{(12)}$O$_2$ layer. Orbital-based differences in
the interface effective-mass are noted by $1/Z_{xy}$=$m^*(xy)/m_{\rm LDA}(xy)$$\sim$1.3, 
in good accordance with Shubnikov-de-Haas measurements~\cite{cav10}, and 
$1/Z_{\tilde{e_g}}$=$m^*(\tilde{e_g})/m_{\rm LDA}(\tilde{e_g})$$\sim$2.1. 
Mass renormalization for Ti$^{(345)}$ $t_{2g}$-like states is rather weak. 
From Tab.~\ref{tab:fillings}, a single electron is associated with each Ti$^{(12)}$ ion
in the VH case. Roughly 3/4 thereof are of stronger-localized $\tilde{e_g}$ and 1/4 
of stronger-itinerant $xy$ character. Significant charge disproportionation between 
Ti$^{(1)}$ and Ti$^{(2)}$ is not observed. The spectral function $A({\bf k},\omega)$ in 
Fig.~\ref{fig5:kresolved}a reveals DMFT self-energy-induced shifts of spectral weight
(see the Appendix) compared to the LDA bands (Fig.~\ref{fig2:dosbands}c). 
\begin{table}[b]
\begin{tabular}{l r|c|c|c|c|c}
 &  & Ti$^{(1)}$    & Ti$^{(2)}$   & Ti$^{(3)}$  & Ti$^{(4)}$ &  Ti$^{(5)}$   \\ \hline
DF   & PM$\;\,$ & 0.02 0.16  & 0.02 0.16 & 0.07 0.05 & 0.07 0.05 & 0.08 0.04  \\ \hline
   & PM$\;\,$   & 0.76 0.24  & 0.79 0.24 & 0.15 0.06 & 0.17 0.06 & 0.24 0.04  \\[0.1cm]
VH & $\uparrow$  & 0.41 0.19  & 0.43 0.19 & 0.07 0.03 & 0.08 0.03 & 0.12 0.02  \\[-0.1cm]
   & \raisebox{1.5ex}{FM} $\downarrow$ & 0.36 0.06 & 0.36 0.07 & 0.07 0.02 & 0.08 0.02 & 0.12 0.02 \\ 
\end{tabular}
\caption{Local Ti$(3d)$ fillings in DF and VH structure from DFT+DMFT. 
DF: Averaged $xz,yz$ and $xy$ values. VH: ($\tilde{e_g}$,$xy$) on 
Ti$^{(12)}$ and ($xz/yz$,$xy$) on Ti$^{(345)}$. \label{tab:fillings}}
\end{table}
A flattened $xy$-like QP band starts right below the Fermi level at $\Gamma$. 
That shifted low-filled $xy$-like QP band carries also Ti$^{(5)}$ weight, i.e. away 
from the direct interface. Therefore lowest-energy QP bands around $\Gamma$
are not only associated with interface-nearest (Ti$^{(12)}$) electrons.  
The hybridized $\tilde{e_g}/xy$ band close to $M'$ rests curt below $\varepsilon_{\rm F}$ 
and the $\tilde{e_g}$- and $xz/yz$-like bands along $\Gamma$$-$$X$ are now partly merged. 
The PM Fermi surface (FS) shown in Fig.~\ref{fig5:kresolved}b displays the experimentally 
detected~\cite{ber13} four-fold star-like shape extending towards $X$.

Albeit numerically demanding, no stable ferromagnetism is deduced for the 
DF structure within DFT+DMFT. Allowing for FM spin polarization in the VH case 
leads to magnetic order, nearly exclusively located in the Ti$^{(12)}$O$_2$ layer 
(cf. Tab.~\ref{tab:fillings}). The FM phase is energetically indeed weakly favored against 
the PM phase with $\Delta E$$\sim$10 meV/Ti$^{(12)}$. A local 
Ti$^{(12)}$ moment of $\sim$0.09$\mu_{\rm B}$ is detected, in excellent agreement with 
experiment~\cite{lee13}. Notably this moment builds up from polarizing both minimal Ti$^{(12)}$
orbitals, yet surprisingly $xy$, though lower filled, has a larger share. 
Figure~\ref{fig6:charge} displays that the real-space spin polarization within the FM phase of 
the VH structure is indeed largest for the $xy$ states.
At $\varepsilon_{\rm F}$, electrons are either of $\tilde{e_g}$ spin-down or
close-to spin-average $t_{2g}$ (only Ti$^{(12)}$ $xy$ contribution shown) flavor
(see Fig.~\ref{fig4:spectrum}b). Thus the FM state in LAO/STO has a truly intriguing 
itinerancy, emerging from many-body scattering between QPs and nearly-localized 
electrons. 
The spectral spin contrast in Fig.~\ref{fig5:kresolved}b shows that
major spin polarization is tied to low-energy. Correlation-induced exchange
splitting leads to a modified fermiology, especially close to $M'$, where a
spin-down QP band with substantial $\tilde{e_g}$ weight finally forms a hole pocket.
While the spin-up FS keeps the star-like shape, the spin-down
FS extends now also towards the neigborhood of $M'$, covering the full BZ.
\begin{figure}[t]
\hspace*{-0.1cm}(a)\hspace*{-0.25cm}\includegraphics*[height=2.95cm]{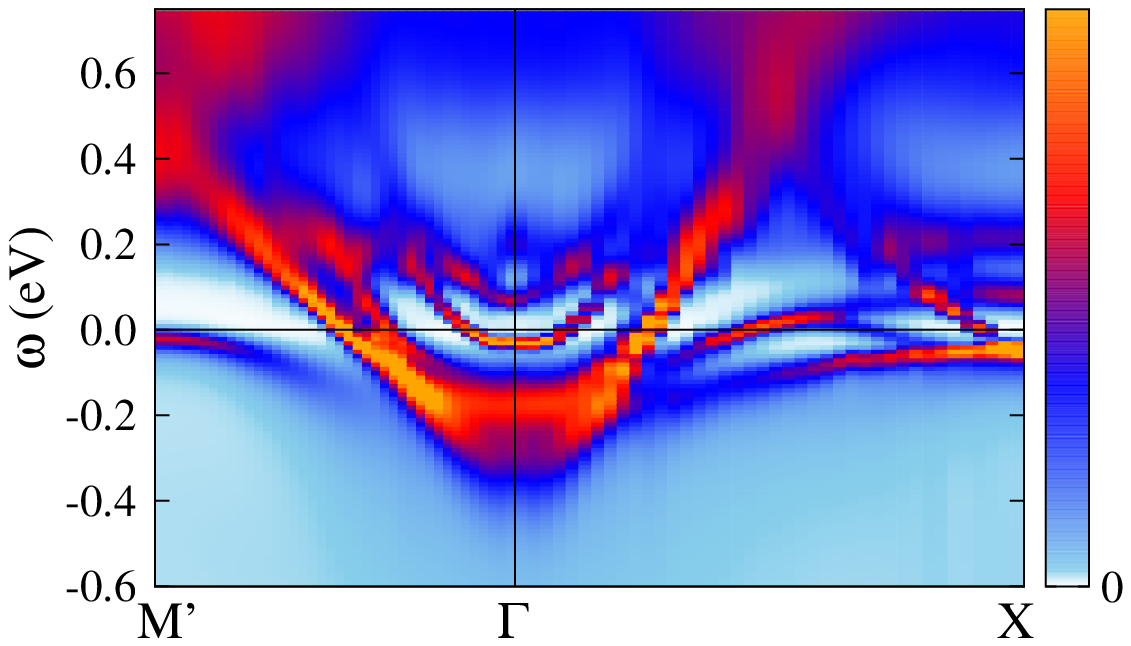}
\hspace*{-0.1cm}\hspace*{-0.0cm}\includegraphics*[height=3.1cm]{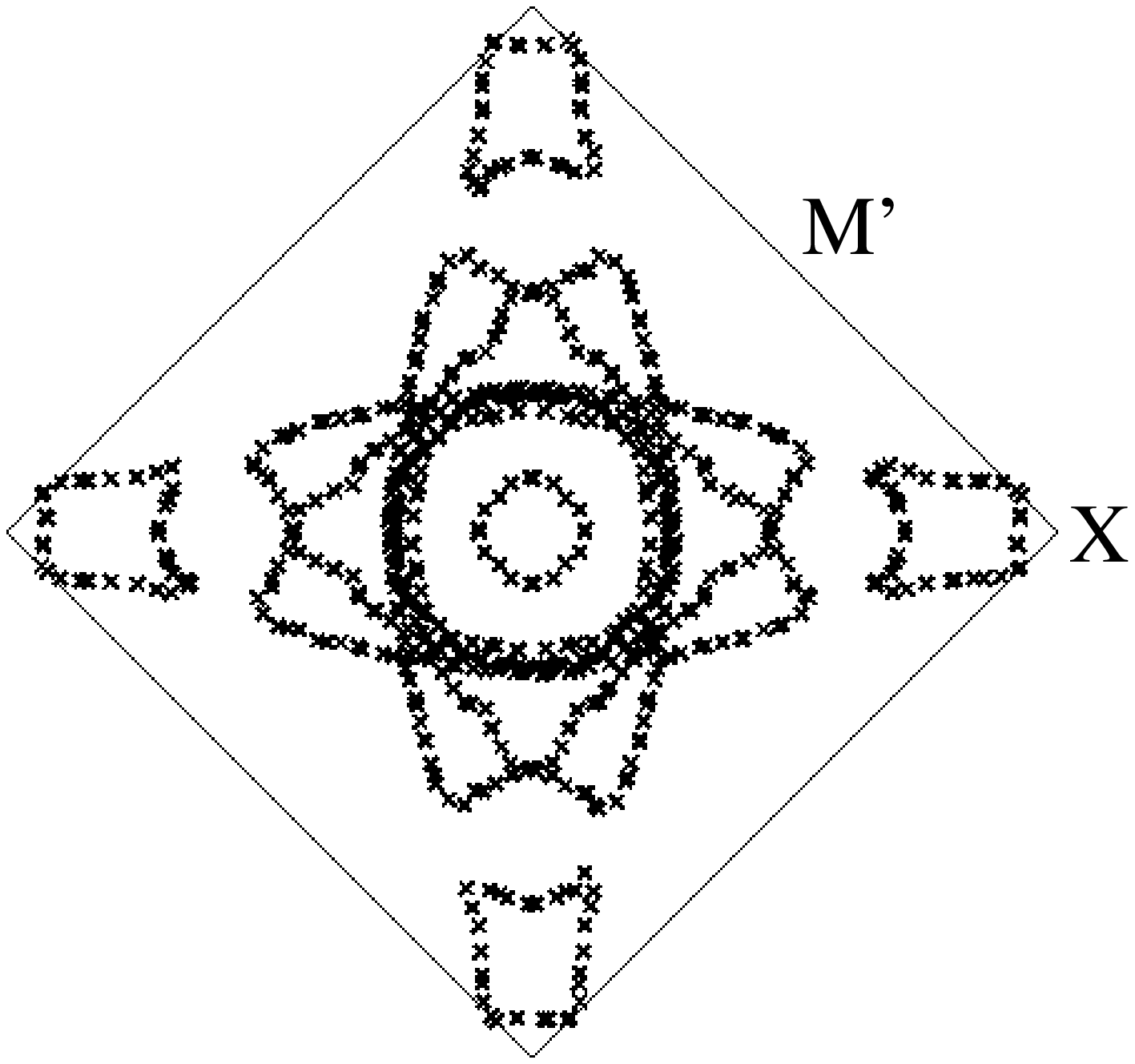}\\[-0.1cm]
\hspace*{-0.1cm}(b)\hspace*{-0.25cm}\includegraphics*[height=2.95cm]{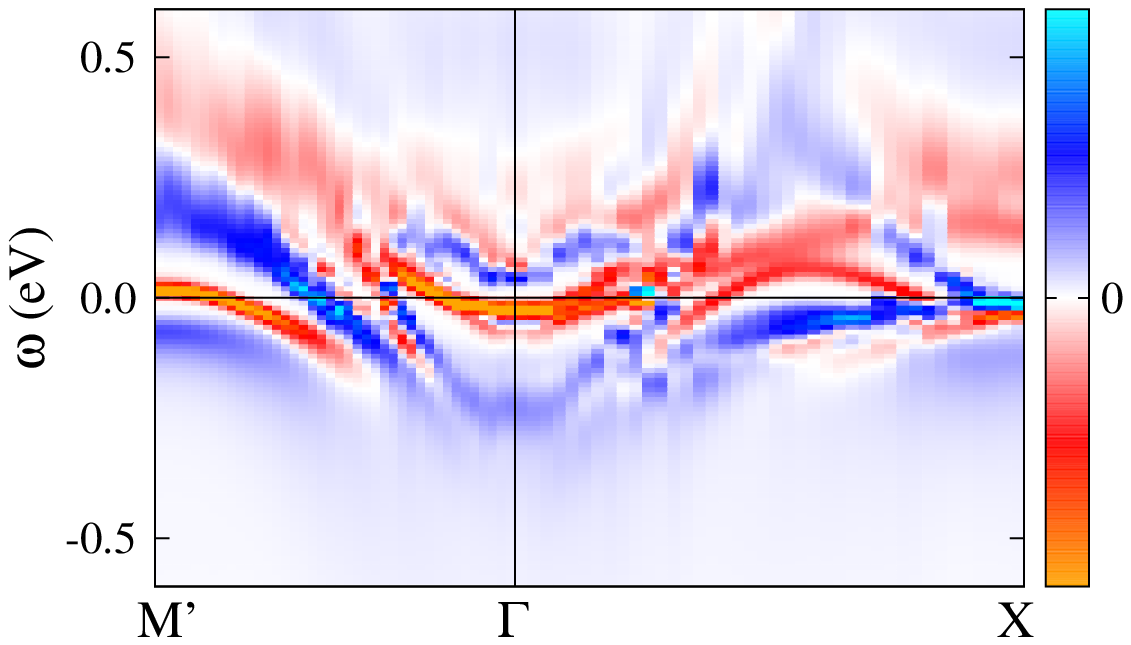}
\hspace*{-0.1cm}\hspace*{-0.0cm}\includegraphics*[height=3.1cm]{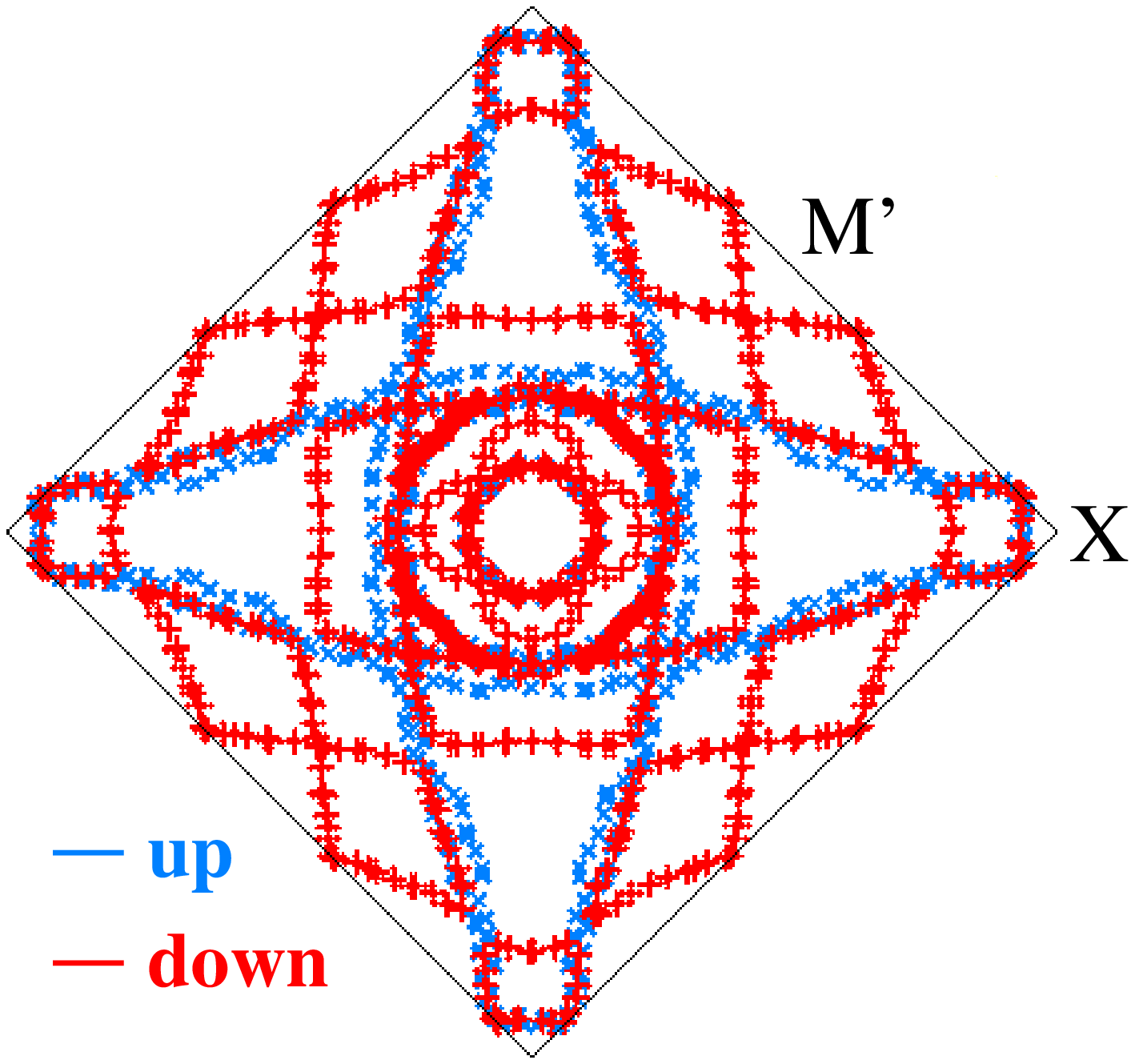}
\caption{(color online) $k$-resolved DFT+DMFT spectral data for VH-LAO/STO.
(a) PM case: $A({\bf k},\omega)$ (left), Fermi surface (right). (b) FM case: 
Spectral spin contrast $A_\uparrow({\bf k},\omega)$$-$$A_\downarrow({\bf k},\omega)$ (left),
spin-resolved Fermi surface (right). \label{fig5:kresolved}}
\end{figure}

Since the PM spectral intensity at $\varepsilon_{\rm F}$ is enhanced with 
correlations, the Stoner criterion rewritten for the Hubbard model as 
$UA(\varepsilon_{\rm F})$$>$1 would be fulfilled. Furthermore it is known that the Hund's
exchange $J_{\rm H}$ triggers itinerant ferromagnetism in the orbital-degenerate 
two-band Hubbard model close to quarter filling~\cite{hel98,mom98}. Especially here due 
to the vacancy-induced minimal scenario of a nearly half-filled $\tilde{e_g}$ orbital 
and a weakly-filled $xy$ orbital, $J_{\rm H}$-driven double-exchange (DE) processes 
dominantly account for FM order~\cite{zen51,and55}. 
However a distinct orbital-character separation into localized spin and 
itinerant electrons as in the standard DE model Hamiltonian is not fully adequate. 
Figure~\ref{fig7:temp}a displays the temperature dependence of the PM Ti$^{(1)}$
self-energy part $\mbox{Im}\,\Sigma(i\omega_{\rm n})$ for Matsubara frequencies 
$\omega_n$=$(2n$$+$$1)\pi T$. Besides showing the noted orbital differentiation in the 
QP renormalization $Z$=$(1-\frac{\partial\mbox{\scriptsize Im}\,\Sigma(i\omega)}{\partial\omega}|_{\omega\rightarrow 0^+})^{-1}$ it reveals insight in the electron-electron
scattering ($\sim$$-Z$$\mbox{Im}\,\Sigma(i0^{+}))$ for $\tilde{e_g}$, $xy$. 
\begin{figure}[t]
\begin{center}
\includegraphics*[height=4.5cm]{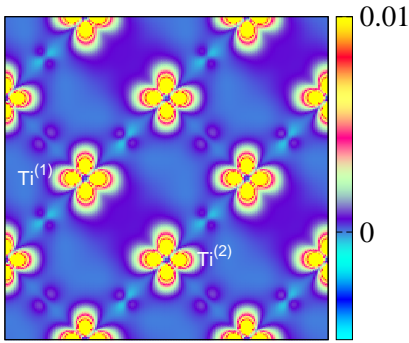}
\end{center}
\caption{(color online)  DFT+DMFT real-space spin contrast 
$\rho_{\uparrow}$$-$$\rho_{\downarrow}$ in the correlated charge density within
Ti$^{(12)}$O$_2$ interface layer for the ferromagnetic phase of the VH 
structure.\label{fig6:charge}}
\end{figure}
\begin{figure}[t]
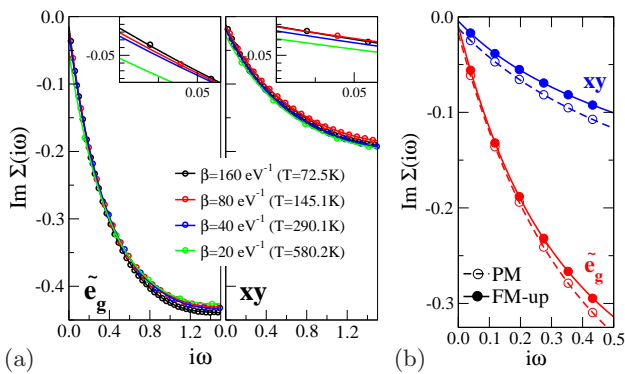

(a)\hspace*{-0.35cm}\includegraphics*[height=4.75cm]{t-imsigma.eps}\hspace*{0.05cm}
(b)\hspace*{-0.35cm}\includegraphics*[height=4.75cm]{spt-imsigma.eps}
\caption{(color online) $\mbox{Im}\,\Sigma(i\omega_{\rm n})$
of Ti$^{(1)}$ impurity self-energy in VH case. Lines are from 5th-order polynomial 
interpolation in a larger frequency range. (a) PM data 
at various $T$ (insets: blow up close to zero frequency). 
(b) Data at $T$=145.1K.\label{fig7:temp}}
\end{figure}
The more localized $\tilde{e_g}$ electrons are less coherent at higher $T$ than 
the $xy$ ones, because of an increased $-\mbox{Im}\,\Sigma(i0^{+})$. 
Consistent with a DE-like mechanism the coherency in the FM phase at low 
temperature is higher than in the PM phase (see Fig.~\ref{fig7:temp}b). Magnetic scattering is 
reduced in the spin-polarized medium due to the satisfaction of the Hund's term. 
From charge self-consistent DFT+DMFT in the FM phase using the spin-{\sl averaged} 
charge density in the DFT part we encounter significantly reduced Ti$^{(12)}$ moments 
$\sim$0.01 $\mu_{\rm B}$ at $T$=290K. In line with experimental findings~\cite{ari11}, 
the Curie temperature of ferromagnetic LAO/STO may thus estimated to lie somewhat above room 
temperature.

\section{Conclusions}
Realistic many-body theory for the LAO/STO interface identifies
the essential ingredients for metallic ferromagnetism and characterizes the underlying
electronic state. Correlations from local Coulomb interactions in the DF interface 
are not strong enough within DFT+DMFT to cause an FM state or straightforward 
charge-ordering instabilities. Ferromagnetism only appears with oxygen vacancies {\sl and}
electronic correlations. Note that pure {\sl static} correlation effects based on 
DFT+U also yield FM moments in the VH case,~\cite{pav12}
but too large in magnitude compared to experiment. Realistic quantum fluctuations
are efficient in reducing the Ti interface moment down to the experimental value of
$\sim$0.1$\mu_{\rm B}$.~\cite{lee13}
The basic many-body behavior is understood within a minimal two-orbital 
$\tilde{e_g}$-$xy$ picture originating in the nearest-interface TiO$_2$ layer, 
derived from the full supercell electronic structure. Strongly 
renormalized temperature-dependent QPs as well as significant incoherent spectral weight
are revealed. In fact the $-1.2$eV peak known from photoemission can be readily
identified as an $\tilde{e_g}$-like lower Hubbard band. Double-exchange
processes between more localized $\tilde{e_g}$ and more itinerant $xy$ lead to
ferromagnetism. The intricate fermiology of especially the FM state involves
at least two ($\tilde{e_g}$,$xy$) carrier types. Details on charge transfers,
orbital contributions, etc. still ask for an inclusion of Ti states from
more distant TiO$_2$ layers. Although we here studied two extreme structural cases, i.e. 
stoichiometric and with high vacancy concentration, the qualitative electronic aspects due to 
oxygen vacancies are believed to hold also for more demanding diluted scenarios. In this regard,
multi-component cluster-expansion based approaches~\cite{lec00} provide a route to access 
the generic thermodynamics of vacancies and other possible defects~\cite{yu14} at the 
LAO/STO interface. Further theoretical work on the correlated electronic structure is needed
to address the magnetic anisotropy and the connection/coexistence of ferromagnetism with 
superconductivity. Finally, time-dependent manipulation of the intricate electronic states 
at the interface via pump-probe techniques may lead to fascinating new non-equilibrium physics.

\begin{acknowledgments}
We are grateful to V. P. Antropov and I. I. Mazin for helpful discussions. 
Calculations were performed at the Juropa Cluster of the J\"ulich Supercomputing Centre 
(JSC) under Project No. hhh08. This research was supported by the DFG-FOR1346 project.
\end{acknowledgments}


\appendix*
\section{Kohn-Sham Hamiltonian and DFT+DMFT Level Shifting} 
There are 10 Ti ions in the supercell. With our choices concerning the respective
correlated subspace, the complete projected Kohn-Sham Hamiltonian has thus 
the dimension 30$\times$30 for the DF calculation and 20$\times$20 for the VH one. 
For instance, the inplane nearest-neighbor (NN) Ti$^{(12)}$ Hamiltonian
block in the latter case reads
\begin{eqnarray}
H_{\tilde{e_g},xy}^{\rm (KS,12,NN)}=&&\nonumber\\
&&\left.\begin{array}{cccc}
\hspace*{-1.5cm}\mbox{Ti}^{(1)} & \hspace*{0.75cm}\mbox{Ti}^{(1)} & \hspace*{1cm}\mbox{Ti}^{(2)} & \hspace*{0.75cm}\mbox{Ti}^{(2)} 
\end{array}\right.\nonumber\\[-0.1cm]
&&\hspace*{-2.5cm}\left(\begin{array}{rrrrrrrr}
  416  &     0 &    -1    &    0   &  -204    &    0    &   -2   &     0   \\
    0  &   729 &     0    &    0   &     0    &  180    &    0   &     0   \\
   -1  &     0 &   416    &    0   &    -2    &    0    &  161   &     0   \\
    0  &     0 &     0    &  729   &     0    &    0    &    0   &  -247   \\
 -204  &     0 &     2    &    0   &   397    &    0    &   -1   &     0   \\
    0  &   180 &     0    &    0   &     0    &  728    &    1   &     0   \\
   -2  &     0 &   161    &    0   &    -1    &    1    &  397   &     0   \\
    0  &     0 &     0    & -247   &     0    &    0    &    0   &   728    
\end{array} \right),
\end{eqnarray}
whereby each onsite Ti subblock consists of a 2$\times$2 matrix for $\tilde{e_g}$ and $xy$
and the values are given in meV.

Besides band renormalization and spectral-weight transfer with correlations, the 
real part $\mbox{Re}\,\Sigma(i0^+)$ of the local self-energy introduces a level shifting.
The additional crystal-field splitting $\Delta_{\rm c}$ due to many-body effects is 
computed as
$\Delta_{\rm c}$=$\varepsilon_{\rm CSC}$$+$$\mbox{Re}\,\Sigma(i0^+)$$-$$\Delta_{\rm DC}$$-$$\varepsilon_{\rm KS}$ for each involved local electronic state individually. Here 
$\varepsilon_{\rm CSC}$ is the converged local level energy from the KS-like part within
DFT+DMFT, $\Delta_{\rm DC}$ is the orbital-independent but site-dependent shift from
the fully-localized double counting and $\varepsilon_{\rm KS}$ the original Kohn-Sham
level energy within LDA. Table~\ref{tab:levels} shows the resulting correlation-induced 
level shifting for the projected orbitals in the VH structural case. Note that the
coherent part of all shifts is of course absorbed in the chemical potential determined
at each DFT+DMFT step.
\begin{table}[h]
\begin{ruledtabular}
\begin{tabular}{c|c|c|c|c|c}
 &  Ti$^{(1)}$ & Ti$^{(2)}$ & Ti$^{(3)}$ & Ti$^{(4)}$  & Ti$^{(5)}$       \\ \hline
$\varepsilon_{\rm KS}$      & 416 729 & 397 728 & 865 1016 & 903 1016 & 870 1086  \\
$\Delta_{\rm c}$            & 244 160 & 249 146 & 191 240  & 200 250  & 279 372 \\
\end{tabular}
\end{ruledtabular}
\caption{Kohn-Sham levels and correlation-induced shifting $\Delta_{\rm c}$ (in meV)
for (left) $\tilde{e_g}$/$(xz/yz)$ and (right) $xy$ at Ti$^{12}$/Ti$^{345}$, in the VH
structure within DFT+DMFT at $T$=145.1K.\label{tab:levels}}
\end{table}
\bibliographystyle{apsrev4-1}
\bibliography{bibextra}

\end{document}